\documentclass[a4paper]{jpconf}

\usepackage[utf8]{inputenc}
\usepackage[T1]{fontenc}
\usepackage{amsmath,amssymb,amsthm}
\usepackage{mathrsfs}  
\usepackage{bbold}

\usepackage{graphicx}

\newcommand{\beq}{\begin{equation}}
\newcommand{\eeq}{\end{equation}}
\newcommand{\beqa}{\begin{eqnarray}}
\newcommand{\eeqa}{\end{eqnarray}}

\begin{document}

\title{Gaussian fluctuations in the two-dimensional BCS-BEC crossover: finite temperature properties}

\author{G Bighin and L Salasnich}

\address{Dipartimento di Fisica e Astronomia ``Galileo Galilei'', 
Universit\`a di Padova, Via Marzolo 8, 35131 Padova, Italy}

\ead{bighin@pd.infn.it}

\begin{abstract}

The role of fluctuations is enhanced in lower dimensionality systems: in a two dimensions off-diagonal long-range order is destroyed by the fluctuations at any finite temperature, drastically modifying the critical properties with respect to the three-dimensional counterpart.
Recently two-dimensional systems of interacting fermions have been the subject of Montecarlo studies and experimental investigations, in particular an ultracold gas of attractive fermions with a widely tunable interaction due to a Feshbach resonance has been realized and the Berezinskii-Kosterlitz-Thouless transition has been observed.

The present work deals with the theoretical description of an ultracold Fermi gas: we discuss the role of Gaussian fluctuations of the order parameter in the equation of state, in particular we take into account the first sound velocity, showing that the inclusion of order parameter fluctuations is needed in order to get the correct composite-boson limit in the strong-coupling regime. The theory is also compared with experimental data. Finally we focus on the superfluid density in the weak-coupling, intermediate and strong-coupling regimes at finite temperature, through which the Berezinskii-Kosterlitz-Thouless critical temperature is obtained.

\end{abstract}

\section{Introduction}

The recent experimental realization of an attractive quasi-two-dimensional ultracold Fermi gas with widely tunable interactions \cite{Martiyanov:2010kt,Frohlich:2011bh,Dyke:2011ja,Sommer:2012bs}, the experimental observation of the critical properties \cite{Murthy:2015ix}, along with many Montecarlo investigations \cite{Bertaina:2011kk,Anderson:2015er} renovated the interested in the two-dimensional BCS-BEC crossover and motivate further theoretical pursuits in this field.

In three dimensions a mean-field theory of the BCS-BEC crossover at $T=0$ can provide good qualitative agreement with experiments; however in lower dimensional systems the role of fluctuations is more important \cite{Hadzibabic:2012} and one expects that a mean-field theory may not provide a reasonable agreement with experimental data. It has been indeed reported that in two dimensions at $T=0$ an ultracold Fermi gas is correctly described by the mean-field theory only in the weak-coupling limit, the correct composite boson equation of state of the strong coupling limit being correctly recovered only when the order parameter fluctuations are taken into account \cite{Salasnich:2015kl}.

In the present work, using a path integral approach \cite{Tempere:2012cm} we derive the one-loop equation of state \cite{He:2015dc}, which we use to calculate the first sound speed and the superfluid density at finite temperature. The first sound speed is compared with experimental data, showing that the Gaussian fluctuations of the order parameter are fundamental even in the zero-temperature limit. The superfluid density as a function of the crossover allows one to determine the Berezinskii-Kosterlitz-Thouless critical temperature \cite{Berezinsky:1970fr,Kosterlitz:1973xp} through the Kosterlitz-Nelson condition \cite{Nelson:1977zz}.

\section{Mean-field results}
Let us consider a two-dimensional system of fermions, interacting through an attractive $s$-wave contact potential, contained in a volume $V=L^2$, described within the grand canonical ensemble at fixed chemical potential $\mu$ and temperature $T$. Within the path integral formalism the partition function for the system can be written as \cite{Tempere:2012cm}
\beq
\mathcal{Z} = \int \mathcal{D} \psi_\sigma \mathcal{D} \bar{\psi}_\sigma e^{- \frac{1}{\hbar} \int_0^{\hbar \beta} \mathrm{d} \tau \int_{L^2} \mathrm{d}^2 r \mathscr{L}}
\label{eq:z}
\eeq
where $\psi$, $\bar{\psi}$ are complex Grassmann fields, $\beta=1/k_B T$, $k_B$ is Boltzmann constant and the (Euclidean) Lagrangian density reads:
\beq
\mathscr{L} = \bar{\psi}_{\sigma} \left[ \hbar \partial_{\tau} 
- \frac{\hbar^2}{2m}\nabla^2 - \mu \right] \psi_{\sigma} 
+ g \, \bar{\psi}_{\uparrow} \, \bar{\psi}_{\downarrow} 
\, \psi_{\downarrow} \, \psi_{\uparrow} \; .
\label{eq:l}
\eeq
The quartic interaction cannot be treated exactly; the Hubbard-Stratonovich \cite{Stratonovich:1958vq,Hubbard:1959ub} transformation introduces an additional auxiliary pairing field, corresponding to a Cooper pair, decoupling the quartic interaction. The transformation essentially boils down to the following identity
\beq
e^{- \frac{1}{\hbar} \int_0^{\hbar \beta} \mathrm{d} \tau \int_{L^2} \mathrm{d}^2 r g \, \bar{\psi}_{\uparrow} \, \bar{\psi}_{\downarrow} 
\, \psi_{\downarrow} \, \psi_{\uparrow}} \propto \int \mathcal{D} \Delta \mathcal{D} \bar{\Delta} \exp^{- \frac{1}{\hbar} S_{hs} [\psi, \bar{\psi}, \Delta, \bar{\Delta}]}
\eeq
where
\beqa
S_{hs} [\psi, \bar{\psi}, \Delta, \bar{\Delta}] = - \int_0^{\hbar \beta} \mathrm{d} \tau \int_{L^2} \mathrm{d}^2 r \left( \frac{|\Delta|^2}{g} + \bar{\Delta} \psi_\downarrow \psi_\uparrow + \Delta  \bar{\psi}_\uparrow \bar{\psi}_\downarrow \right) \; ,
\eeqa
which can be straightforwardly verified by completing the square and performing the Gaussian integration. The partition function now reads:
\beq
\mathcal{Z} = \int \mathcal{D} \psi_\sigma \mathcal{D} \bar{\psi}_\sigma \mathcal{D} \Delta \mathcal{D} \bar{\Delta} e^{- \frac{1}{\hbar} \int_0^{\hbar \beta} \mathrm{d} \tau \int_{L^2} \mathrm{d}^2 r \mathscr{L}}
\label{eq:z2}
\eeq
with the following Lagrangian density
\beq
\mathscr{L} =
\bar{\psi}_{\sigma} \left[  \hbar \partial_{\tau} 
- {\hbar^2\over 2m}\nabla^2 - \mu \right] \psi_{\sigma} 
+ \bar{\Delta} \, \psi_{\downarrow} \, \psi_{\uparrow} 
+ \Delta \bar{\psi}_{\uparrow} \, \bar{\psi}_{\downarrow} 
- {|\Delta|^2\over g} \; .
\label{ltilde}
\eeq
The integration over $\psi$, $\bar{\psi}$ can be carried out exactly \cite{Tempere:2012cm}, obtaining:
\beq
\mathcal{Z} = \int \mathcal{D} \Delta \mathcal{D} \bar{\Delta} e^{- \frac{1}{\hbar} \int_0^{\hbar \beta} \mathrm{d} \tau \int_{L^2} \mathrm{d}^2 r \left( - \ln ( - \mathbb{G}^{-1} \right) - {|\Delta|^2 / g})}
\label{eq:z3}
\eeq
and the physics of the system is encoded in the inverse Green's function, which in coordinate-space representation reads:
\beq
- \mathbb{G}^{-1} = \begin{pmatrix} \hbar \partial_\tau - \frac{\hbar^2}{2m} \nabla^2 - \mu & \Delta(\mathbf{x},\tau) \\
\bar{\Delta}(\mathbf{x},\tau) & \hbar \partial_\tau + \frac{\hbar^2}{2m} \nabla^2 +\mu \end{pmatrix} \; .
\eeq
The full inverse Green function can be decomposed in a mean-field component $- \mathbb{G}^{-1}_0$, where the pairing field $\Delta$ is replaced by its uniform and constant saddle point value, plus a fluctuation part $\mathbb{F}$:
\beq
- \mathbb{G}^{-1} =  - \mathbb{G}^{-1}_{0} + \mathbb{F} = \begin{pmatrix} \hbar \partial_\tau - \frac{\hbar^2}{2m} \nabla^2 - \mu & \Delta_0 \\
\Delta_0 & \hbar \partial_\tau + \frac{\hbar^2}{2m} \nabla^2 +\mu \end{pmatrix} + \begin{pmatrix} 0 & \eta(\mathbf{x},\tau) \\
\bar{\eta}(\mathbf{x},\tau) & 0 \end{pmatrix}
\label{eq:expansion}
\eeq
and clearly $\Delta(\mathbf{x},\tau) = \Delta_0 + \eta(\mathbf{x},\tau)$. We now analyze the mean-field approximation, which simply consists in neglecting the fluctuation field $\eta$ and, consequently, $\mathbb{F}$. The single-particle excitation spectrum is found solving for the poles of the Nambu-Gor'kov Green's function $\mathbb{G}_0$ in momentum space \cite{Stoof:2009tf}:
\beq
E_\mathbf{k} = \sqrt{(k^2/2m - \mu)^2 + \Delta_0^2} \;.
\label{eq:ek}
\eeq
Using the bound state equation to relate the interaction strength $g$ to the bound state\footnote{A bound state is present in two-dimension for every value of the attractive interaction.} energy $\epsilon_B$
\beq 
- \frac{1}{g} = \frac{1}{2L^2} \sum_{\bf k} \frac{1}{\epsilon_k + 
\frac{1}{2} \epsilon_B}
\label{g-eb}
\eeq
at $T=0$ the integrations defining $\mathcal{Z}$ can be carried out analytically in two dimensions,  finally the thermodynamic grand potential reads:
\beqa 
\Omega_{mf} (\mu, \Delta_0) = - \frac{1}{\beta} \ln \mathcal{Z} =  - {m L^2\over 4\pi \hbar^2} \Big[ 
\mu^2 + \mu \sqrt{\mu^2+\Delta_0^2} + {1\over 2} \Delta_0^2 - \Delta_0^2 \ln{\Big({-\mu + \sqrt{\mu^2 + \Delta_0^2} 
\over \epsilon_B}\Big)} \Big]  \; . 
\label{eq:mfeosa}
\eeqa
From the grand potential we impose the saddle-point condition for $\Delta_0$,
i.e. $(\partial \Omega_{mf} / \partial \Delta_0)_{\mu,V} = 0$, obtaining the gap equation:
\beq
\Delta_0 = \sqrt{2\epsilon_b ( \mu + {1\over 2}\epsilon_B ) } \; , 
\label{gapeq}
\eeq 
inserting the result from Eq. (\ref{gapeq}) into Eq. (\ref{eq:mfeosa}) one gets the mean-field equation of state \cite{Marini:1998ej}:
\beq 
\Omega_{mf} (\mu) = - {m L^2\over 2\pi \hbar^2} 
(\mu + {1\over 2} \epsilon_B )^2  \; . 
\label{eq:mfeos}
\eeq
The number of particles is obtained from the thermodynamic relation $N = - \partial \Omega / \partial \mu$ and reads:
\beq
N =  {m L^2\over \pi \hbar^2} 
(\mu + {1\over 2} \epsilon_B ) \; .
\eeq

\section{Pairing fluctuations and equation of state}
\begin{figure}[h!]
\centering
\includegraphics[width=0.6\linewidth]{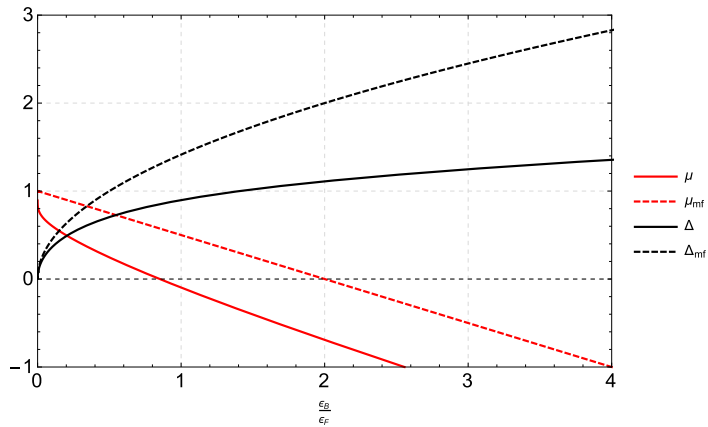}\hspace{2pc}%
\caption{The chemical potential $\mu$ (red lines) and the pairing gap $\Delta_0$ (black lines), as calculated from the mean-field equation of state (dashed lines) and from the one-loop equation of state (solid lines), as a function of the scaled binding energy $\epsilon_B/\epsilon_F$.}
\label{fig:mudelta}
\end{figure}

In the previous Section we simply discarded the fluctuation part of the inverse Green's function, here we rewrite the logarithm appearing in Eq. (\ref{eq:z3}) as
\beq
\ln \left( - \mathbb{G}^{-1} \right) = \ln \left(  - \mathbb{G}^{-1}_{0} \left( \mathbb{1} - \mathbb{G}^{-1}_{0} \mathbb{F} \right) \right) = \ln \left(  - \mathbb{G}^{-1}_{0} \right) + \ln \left( \mathbb{1} - \mathbb{G}^{-1}_{0} \mathbb{F} \right)
\label{eq:f1}
\eeq
and the Gaussian (one-loop) approximation consists in the following expansion for the second term in the r.h.s. of Eq. (\ref{eq:f1})
\beq
\ln \left( \mathbb{1} - \mathbb{G}^{-1}_{0} \mathbb{F} \right) = \sum_{m=1}^\infty \frac{\left( \mathbb{G}^{-1}_{0} \mathbb{F} \right)^m}{m} \approx - \mathbb{G}^{-1}_{0} \mathbb{F}  - \frac{1}{2} \mathbb{G}^{-1}_{0} \mathbb{F}  \mathbb{G}^{-1}_{0} \mathbb{F} \; .
\eeq
The pairing field has been written as $\Delta(\mathbf{x},\tau) = \Delta_0 + \eta (\mathbf{x},\tau)$ where $\Delta_0$ the saddle point value, so the fluctuation fields $\eta$, $\bar{\eta}$ are small, justifying the present expansion in powers of $\mathbb{F}$. Re-arranging the terms and rewriting the action in momentum space \cite{Tempere:2012cm} one gets
\beq
\mathcal{Z} = \int \mathcal{D} \eta \mathcal{D} \bar{\eta} e^{- \frac{1}{\hbar} S_g [\eta,\bar{\eta}]}
\label{eq:z4}
\eeq
with the following action
\beq
S_{g} [\eta,\bar{\eta}] = {1\over 2} \sum_{\mathbf{q}, m} 
({\bar\eta}(\mathbf{q}, \mathrm{i} \Omega_m),\eta(-\mathbf{q}, - \mathrm{i} \Omega_m)) \ \mathbb{M} (\mathbf{q}, \mathrm{i} \Omega_m) \left(
\begin{array}{c}
\eta(\mathbf{q}, \mathrm{i} \Omega_m) \\ 
{\bar\eta}(-\mathbf{q}, - \mathrm{i} \Omega_m) 
\end{array}
\right) \; ,
\eeq
having introduced the Fourier transform of the fluctuation field $\eta$, along with the bosonic Matsubara frequencies $\Omega_m = \left( 2 m + 1 \right) \pi/ \beta$; the matrix elements of the inverse pair fluctuation propagator $\mathbb{M}$, along with a complete derivation can be found in Ref. \cite{Tempere:2012cm}. The poles of the propagator determine the collective mode spectrum, which in the strong-coupling regime can be parametrized as
\beq
\hbar \omega_q = \sqrt{\epsilon_q \left( \lambda \epsilon_q + 2 m c_s^2 \right)} \; ,
\eeq
where $\lambda$ and $c_s$ are a function of the crossover. In order to write the one-loop contribution to the equation of state \cite{Diener:2008kh,He:2015dc} one needs to introduce a regularization procedure \cite{Diener:2008kh,He:2015dc}, as follows:
\beq
\Omega_g (\mu) = {1\over 2\beta} \sum_{\mathbf{q}, m} \ln \left[ \frac{\mathbb{M}_{11} (\mathbb{q}, \mathrm{i} \Omega_m)}{\mathbb{M}_{22} (\mathbf{q}, \mathrm{i} \Omega_m)} \det(\mathbb{M} (\mathbf{q}, \mathrm{i} \Omega_m)) \right] e^{\mathrm{i} \Omega_m 0^+} \;.
\label{eq:fulleos}
\eeq
In Fig. \ref{fig:mudelta} we plot the mean-field value of the chemical potential and of the pairing gap as a function of the crossover, calculated using the mean-field equation of state in Eq. (\ref{eq:mfeos}) and the full one-loop equation of state given by the sum of Eq. (\ref{eq:mfeos}) and Eq. (\ref{eq:fulleos}).

\section{First sound}

\begin{figure}[h]
\begin{minipage}{0.48\linewidth}
\includegraphics[width=0.95\linewidth]{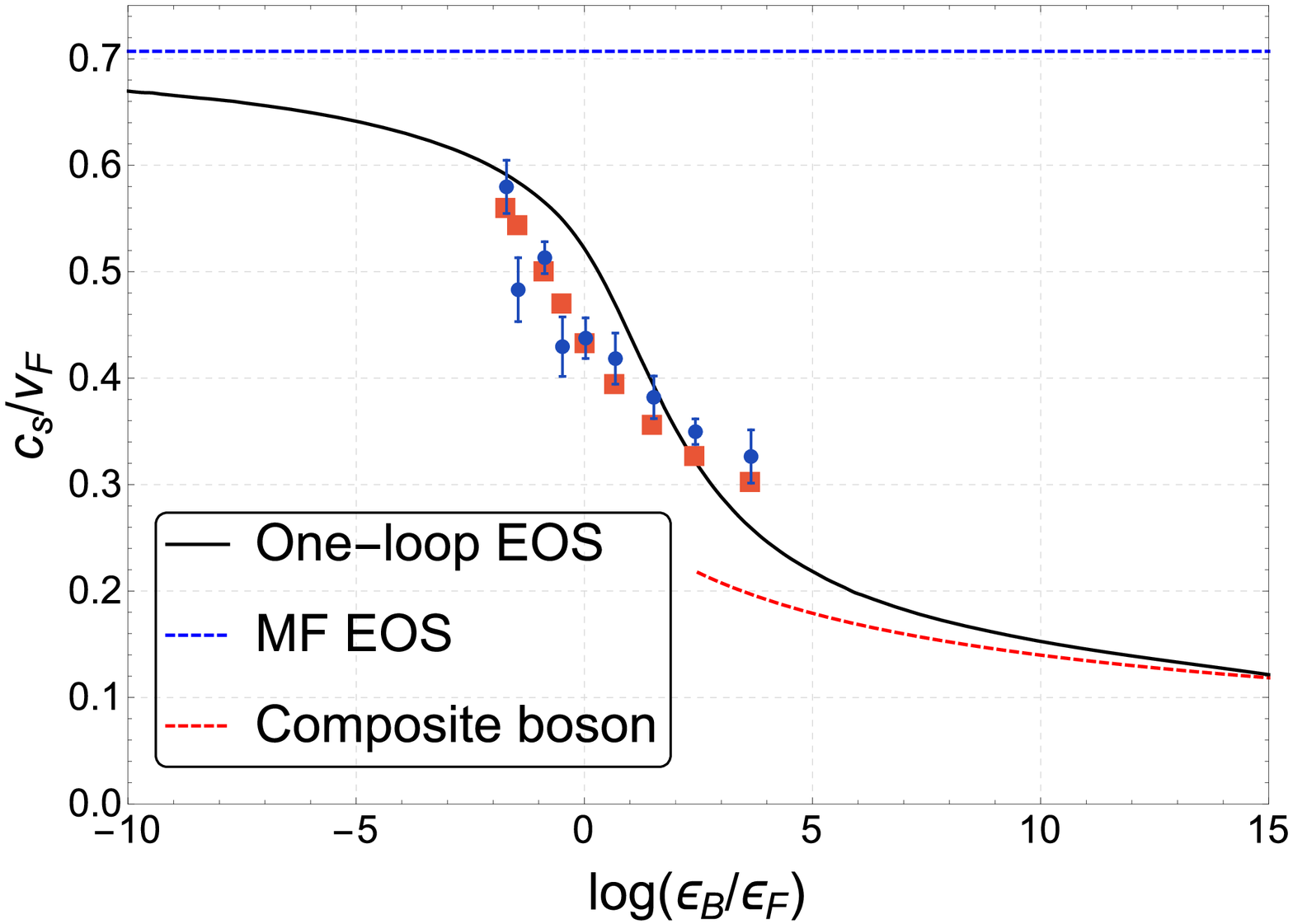}
\caption{\label{fig:cs}The first sound speed at $T=0$ from the one-loop equation of state (solid black line), compared with the theoretical prediction using the mean-field equation of state (blue dashed line) and the composite boson limit valid for $\epsilon_B \gg \epsilon_F$ (red dashed line). The data points are preliminary experimental data reported in \cite{Luick:2014}, blue dots for data measured from the speed of density waves and red squares for data measured from the equation of state.}
\end{minipage}\hspace{0.04\linewidth}%
\begin{minipage}{0.48\linewidth}
\includegraphics[width=0.95\linewidth]{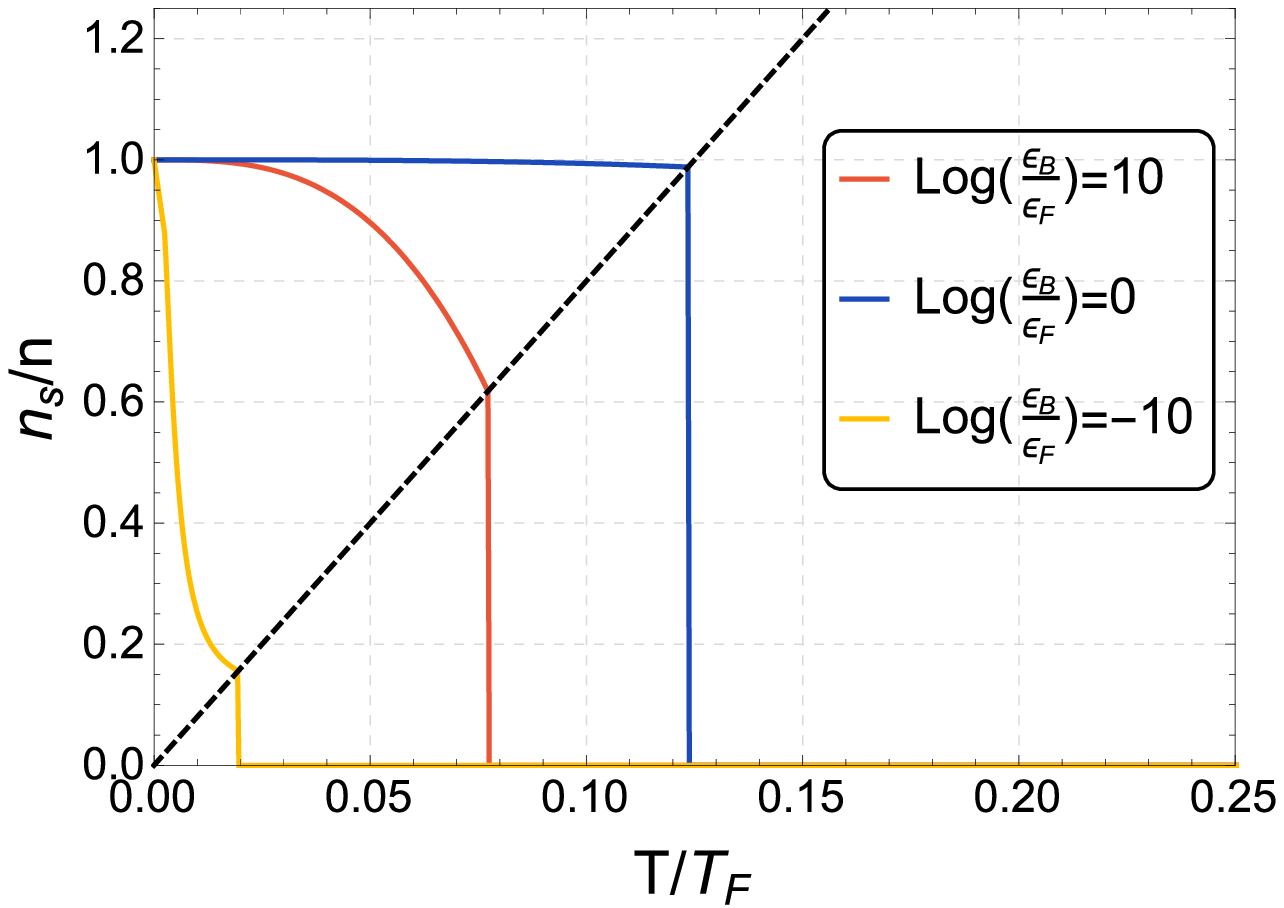}
\caption{\label{fig:rhos}Superfluid density for three different values of the scaled binding energy, from the weakly-coupled regime to the strong interacting one. The black dashed line marks the Nelson-Kosterlitz condition, setting the Berezinskii-Kosterlitz-Thouless critical temperature $T_{BKT}$; a more detailed analysis of $T_{BKT}$ across the whole crossover is reported in \cite{Bighin:2016uk}.}
\end{minipage} 
\end{figure}

Within this framework the first sound speed at $T=0$ can be calculated using the elegant thermodynamic relation \cite{Lipparini:2008} due to Landau
\beq
c_s^2 = \frac{n}{m} \frac{\partial \mu}{\partial n} \; .
\eeq
Using the mean-field equation of state in Eq. (\ref{eq:mfeos}) one finds the result
\beq
c_s^2 = \frac{v_f^2}{2} \; ,
\eeq
where $v_F$ is the Fermi velocity, this result being plotted in dashed blue in Fig. \ref{fig:cs}. On the other hand, introducing the one-loop equation of state in Eq. (\ref{eq:fulleos}) one finds that the mean-field result is accurate only in the deep-BCS regime: when the coupling is increased the sound speed greatly deviates from its mean-field value tending to the composite boson limit \cite{Salasnich:2015kl}
\beq
c_s^2 = \frac{4 \pi \hbar^2}{m^2_B} \frac{n_B}{\ln \left( \frac{1}{n_B a_B^2} \right)}
\eeq
shown in dashed red in Fig. \ref{fig:cs}, $m_B=2 m$ being the mass of each composite boson, $n_B = n / 2$ being the bosonic density and $a_B = 0.551 a_s$ being the boson-boson scattering length \cite{Salasnich:2015kl}. Our theoretical findings are also compared with preliminary experimental data \cite{Luick:2014}, showing good agreement, see Fig \ref{fig:cs}. The results can be readily extended to finite temperature \cite{Bighin:2016uk} using the thermodynamic approach in \cite{Salasnich:2010jw}, however the temperature dependance is very weak across the whole crossover between $T=0$ and the Berezinskii-Kosterlitz-Thouless critical temperature $T_{BKT}$ \cite{Bighin:2016uk}.

We stress that the present result in Fig. \ref{fig:cs} clearly shows that the inclusion of Gaussian fluctuations in the equation of state is of fundamental importance in the two-dimensional BCS-BEC crossover, the mean-field equation of state being accurate only in the deep-BEC regime.

\section{Superfluid density}

The fermionic and bosonic contributions to the normal density are calculated using Landau's quasiparticle excitation formulas \cite{Fetter:1971}
\beq
n_{n,f} = \beta \int \frac{\mathrm{d}^2 k}{(2 \pi)^2} k^2 \frac{e^{\beta E_k}}{(e^{\beta E_k} + 1)^2}
\label{eq:nnf}
\eeq
and
\beq
n_{n,b} = \frac{\beta}{2} \int \frac{\mathrm{d}^2 q}{(2 \pi)^2} q^2 \frac{e^{\beta \omega_q}}{(e^{\beta \omega_q} - 1)^2} \;.
\label{eq:nnb}
\eeq
The superfluid density is then $n_s=n-n_{n,f}-n_{n,b}$, having neglected the hybridization between the single-particle and collective excitations which gets relevant at higher temperature scales \cite{Taylor:2006ky,Bighin:2016uk}. As mentioned in the Introduction the role of the fluctuations is enhanced in lower dimensionalities \cite{Hadzibabic:2012}, in particular in one- and two-dimensional systems there cannot the spontaneous breaking of a continuos symmetry at finite temperature \cite{Mermin:1966da,Coleman:1973ci}; as a consequence off-diagonal long-range order is achieved strictly only at $T=0$. Nonetheless in a two-dimensional systems a finite-temperature transition is to be observed, between the normal state and a state characterized by \textit{algebraic} off-diagonal long-range order \cite{Berezinsky:1970fr,Kosterlitz:1973xp}. Figure \ref{fig:rhos} shows three superfluid density profiles for $\log (\epsilon_B/\epsilon_F)=10$, $\log (\epsilon_B/\epsilon_F)=0$ and $\log (\epsilon_B/\epsilon_F)=-10$, respectively in the deep-BEC, intermediate and deep-BCS regimes, along with the line corresponding to the Nelson-Kosterlitz condition \cite{Nelson:1977zz}
\beq
k_B T_{BKT} = \frac{\hbar^2 \pi}{8 m} n_s (T_{BKT})
\label{eq:kncondition}
\eeq
determining the critical temperature $T_{BKT}$.

With respect to the mean-field result \cite{Salasnich:2013kq} a full Gaussian-level equation of state here predicts a lower critical temperature in the deep-BEC regime. This treatment can be extended to the whole crossover, a full comparison with the recent experimental observation of $T_{BKT}$ \cite{Murthy:2015ix} in a quasi-two-dimensional ultracold Fermi gas is reported in Ref. \cite{Bighin:2016uk}.

\section{Conclusions}

The BCS-BEC crossover in a two-dimensional Fermi gas is the subject of great interest, both from an experimental and theoretical point of view. In this work, using a one-loop equation of state, we have calculated the first sound speed and the superfluid density, showing that the Gaussian fluctuations are needed in order to recover the correct composite boson limit even in the zero-temperature case. We also discussed the superfluid density at finite temperature, using this result to calculate the Berezinskii-Kosterlitz-Thouless critical temperature \cite{Bighin:2016uk}.

The treatment in this work can easily extended to account for the description of many properties of two-dimensional ultracold Fermi gases, e.g. the second sound, the Berezinskii-Kosterlitz-Thouless critical temperature across the whole crossover, the condensate fraction \cite{Salasnich:2007im,Bighin:2016uk}. Moreover the energy density, the pressure and the contact have been calculated \cite{He:2015dc} within an approach equivalent to the present one and show good agreement with Montecarlo data.

A description of a trapped non-uniform system, as well as a full calculation of the superfluid density beyond the approximation in Eqs. (\ref{eq:nnb}) and (\ref{eq:nnf}) is the subject of ongoing work.

\section*{Acknowledgements}

The authors gratefully acknowledge Ministero Istruzione Universit\`a Ricerca (PRIN project 2010LLKJBX) for partial support.


\section*{References}

\begin{thebibliography}{99}

\bibitem{Martiyanov:2010kt}
Martiyanov K, Makhalov V and Turlapov A 2010 {\em Phys. Rev. Lett.\/}
  {\bf 105} 030404--4

\bibitem{Frohlich:2011bh}
Fr{\"o}hlich B, Feld M, Vogt E, Koschorreck M, Zwerger W and K{\"o}hl M 2011
  {\em Phys. Rev. Lett.\/} {\bf 106} 105301--4

\bibitem{Dyke:2011ja}
Dyke P, Kuhnle E~D, Whitlock S, Hu H, Mark M, Hoinka S, Lingham M, Hannaford P
  and Vale C~J 2011 {\em Phys. Rev. Lett.\/} {\bf 106} 105304--4

\bibitem{Sommer:2012bs}
Sommer A~T, Cheuk L~W, Ku M~J~H, Bakr W~S and Zwierlein M~W 2012 {\em Physical
  Review Letters\/} {\bf 108} 045302--5

\bibitem{Murthy:2015ix}
Murthy P~A, Boettcher I, Bayha L, Holzmann M, Kedar D, Neidig M, Ries M~G, Wenz
  A~N, Z{\"u}rn G and Jochim S 2015 {\em Phys. Rev. Lett.\/} {\bf 115}
  010401--6

\bibitem{Bertaina:2011kk}
Bertaina G and Giorgini S 2011 {\em Phys. Rev. Lett.\/} {\bf 106}
  110403--4

\bibitem{Anderson:2015er}
Anderson E~R and Drut J~E 2015 {\em Phys. Rev. Lett.\/} {\bf 115}
  115301--5

\bibitem{Hadzibabic:2012}
Hadzibabic Z and K\"ohl M 2012 {\em {Ultracold Bosonic and Fermionic Gases}\/}, vol. 5 (Elsevier) pp 95-120

\bibitem{Salasnich:2015kl}
Salasnich L and Toigo F 2015 {\em Phys. Rev. A\/} {\bf 91} 011604--5

\bibitem{Tempere:2012cm}
Tempere J and Devreese J~P~A 2012 {\em {Superconductors - Materials, Properties
  and Applications}\/} (InTech, Rijeka) pp 1--32

\bibitem{He:2015dc}
He L, L{\"u} H, Cao G, Hu H and Liu X~J 2015 {\em Phys. Rev. A\/} {\bf 92}
  023620--15

\bibitem{Berezinsky:1970fr}
Berenzinskii V~L 1972 {\em Soviet Physics JETP\/} {\bf 34}

\bibitem{Kosterlitz:1973xp}
Kosterlitz J~M and Thouless D~J 1973 {\em Journal of Physics C: Solid State
  Physics\/} {\bf 6} 1181--1203

\bibitem{Nelson:1977zz}
Nelson D~R and Kosterlitz J~M 1977 {\em Phys. Rev. Lett.\/} {\bf 39}
  1201--1205

\bibitem{Stratonovich:1958vq}
Stratonovich R~L 1958 {\em Sov. Phys. Doklady\/} {\bf 2} 416--419

\bibitem{Hubbard:1959ub}
Hubbard J 1959 {\em Phys. Rev. Lett.\/} {\bf 3} 77--80

\bibitem{Stoof:2009tf}
Stoof H~T~C, Gubbels K~B and Dickerscheid D~B~M 2009 {\em {Ultracold Quantum
  Fields}\/} (Dordrecht: Springer)

\bibitem{Marini:1998ej}
Marini M, Pistolesi F and Strinati G~C 1998 {\em Eur. Phys. J. B\/} {\bf 1}
  151--159

\bibitem{Diener:2008kh}
Diener R~B, Sensarma R and Randeria M 2008 {\em Phys. Rev. A\/} {\bf 77}
  023626--21

\bibitem{Lipparini:2008}
Lipparini~E 2008 {\em {Modern Many-Particle Physics}\/}, 2nd ed. (World Scientific, Singapore)

\bibitem{Luick:2014} Luick N 2014 MSc Thesis (University of Hamburg)

\bibitem{Bighin:2016uk}
Bighin G and Salasnich L 2016 {\em Phys. Rev. B\/} {\bf 93} 014519

\bibitem{Salasnich:2010jw}
Salasnich L 2010 {\em Phys. Rev. A\/} {\bf 82} 063619--7

\bibitem{Fetter:1971}
Fetter A~L and Walecka J~D 1971 {\em {Quantum Theory of Many-Particle Systems
  (Dover Books on Physics)}\/} (Dover Publications)

\bibitem{Taylor:2006ky}
Taylor E, Griffin A, Fukushima N and Ohashi Y 2006 {\em Phys. Rev. A\/}
  {\bf 74} 063626--15

\bibitem{Mermin:1966da}
Mermin N~D and Wagner H 1966 {\em Phys. Rev. Lett.\/} {\bf 17}
  1133--1136

\bibitem{Coleman:1973ci}
Coleman S~R 1973 {\em Commun. Math. Phys.\/} {\bf 31} 259--264

\bibitem{Salasnich:2013kq}
Salasnich L, Marchetti P~A and Toigo F 2013 {\em Phys. Rev. A\/} {\bf 88}
  053612--7

\bibitem{Salasnich:2007im}
Salasnich L 2007 {\em Phys. Rev. A\/} {\bf 76} 015601--4

\end{thebibliography}
\end{document}